\documentstyle[emulateapj]{article}
\begin{document}

\title{A Shock-Induced Pair of Superbubbles in the High-Redshift
        Powerful Radio Galaxy MRC 0406$-$244
\footnote{Based on data collected at Subaru Telescope, which is
        operated by the National Astronomical Observatory of Japan.}
}

\author{Yoshiaki Taniguchi$^1$, Youichi Ohyama$^2$,
Takashi Murayama$^1$, Michitoshi Yoshida$^3$, Nobunari Kashikawa$^4$,
Masanori Iye$^4$, Kentaro Aoki$^5$, Toshiyuki Sasaki$^2$, George Kosugi$^2$,
Tadafumi Takata$^2$, Yoshihiko Saito$^4$, Koji S. Kawabata$^4$,
Kazuhiro Sekiguchi$^2$, Kiichi Okita$^2$, Yasuhiro Shimizu$^3$,
Motoko Inata$^4$, Noboru Ebizuka$^6$, Tomohiko Ozawa$^7$,
Yasushi Yadomaru$^7$, Hiroko Taguchi$^8$, Yasuhiro Shioya$^1$,
Shingo Nishiura$^{1, 9}$, Hiroshi Sudou$^1$, Tohru Nagao$^1$,
Saeko Noda$^1$, Yuko Kakazu$^1$, Masaru Ajiki$^1$, Shinobu S. Fujita$^1$,
\& Rie R. Kobayashi$^1$}

\affil{$^1$Astronomical Institute, Graduate School of Science,
            Tohoku University, Aramaki, Aoba, Sendai 980-8578, Japan \\
        $^2$Subaru Telescope Office, National Astronomical Observatory 
            of Japan, 650 N. A`ohoku Place, University Park,
            Hilo, HI 96720, USA \\
        $^3$Okayama Astrophysical Observatory, National Astronomical
	Observatory of Japan, Kamogata-cho, Asakuchi-gun, Okayama 719-0232,
            Japan \\
        $^4$National Astronomical Observatory of Japan, 2-21-1 Osawa,
            Mitaka, Tokyo 181-8588, Japan \\
        $^5$Japan Science and Technology Corporation, Tokyo 102-0081, Japan \\
	$^6$Communications Research Laboratory, 4-2-1 Nukui-Kitamachi,
	    Koganei, Tokyo 184-8795, Japan \\
	$^7$Misato Observatory, 180 Matsugamine, Misato-cho, Amakusa-gun,
	    Wakayama, 640-1366, Japan\\
	$^8$Department of Astronomy and Earth Sciences, Tokyo Gakugei
	    University, 4-1-1 Nukui-Kitamachi, Koganei, Tokyo 184-8501, Japan \\
        $^9$Kiso Observatory, University of Tokyo, Mitake, Kiso, Nagano
            397-0101, Japan}

\begin{abstract}
We present new optical spectroscopy of the high-redshift powerful
radio galaxy MRC 0406$-$244 at redshift of 2.429.
We find that the two extensions toward NW and SE probed in the 
rest-frame ultraviolet image are heated mainly by the nonthermal
continuum of the active galactic nucleus.
However, each extension shows a shell-like morphology, suggesting that
they are a pair of superbubbles induced by the superwind activity
rather than by the interaction between the radio jet and the ambient
gas clouds. If this is the case, the intense starburst
responsible for the formation of superbubbles could occur
$\sim 1 \times 10^9$ yr ago. On the other hand, the age of the radio
jets may be of the order of $\sim 10^6$ yr, being much shorter
than the starburst age. Therefore, the two events, i.e., the starburst
and the radio-jet activities, are independent phenomena. However,
their directions of the expanding motions could be governed by the rotational
motion of the gaseous component in the host galaxy. This idea
appears to explain the alignment effect of MRC 0406$-$244.
\end{abstract}


\section{Introduction}

Since high-redshift ($z$) powerful radio galaxies (HzPRGs) provide us 
a unique opportunity to investigate the formation and evolution 
of both galaxies and active galactic nuclei (AGN), follow-up investigations
have been made intensively for these past two decades
(Chambers, Miley, \& van Breugel 1990; Eales \& Rawlings 1993, 1996;
R\"ottgering et al. 1995; Best et al. 1996; see for a review McCarthy 1993).
One of important problems related to HzPRGs is the so-called
^^ ^^ alignment effect"; there is the strong correlation between
the position angles of the radio axis and the rest-frame optical
and ultraviolet (UV) continua for PRGs at redshifts above $\sim$ 0.6
(Chambers, Miley, \& van Breugel
1987; McCarthy et al. 1987; see also Djorgovski et al. 1987).
The origin of this alignment effect has been in debate 
(e.g., McCarthy 1993) and there are two favorable ideas.
One is the jet-induced star formation in which the UV continuum
emission is considered to arise from massive stars formed in
the shocked region (e.g., Chambers et al. 1987; McCarthy et al. 
1987; Rees 1989; Begelman \& Cioffi 1989; Daly 1990). 
The kinematics and ionization state of HzPRGs often suggest the importance
of the shock heating (e.g., Best et al. 2000; De Breuck et al. 2000).
An alternative idea is the scattering of
an anisotropic continuum radiation from a central engine of AGN
(Tadhunter, Fosbury, \& di Serego Alighieri 1988; Fabian 1988;
di Serego Alighieri et al. 1989; Scarrott, Rolph, \& Tadhunter 1990;
Cimatti et al. 1994).  The strong linear polarization in the rest-frame
UV continua has been considered as strong evidence for this idea 
(e.g., Cimatti et al. 1998 and references therein).

Among the HzPRGs with evidence for the alignment effect,
MRC 0406$-$244 has been investigated in detail
(McCarthy et al. 1992; Eales \& Rawlings 1993, 1996; Rush et al. 1997).
This radio galaxy was discovered in the MRC/1Jy radio source survey
and identified as a HzPRG at $z$ = 2.429 (McCarthy et al. 1991, 1996).
It shows core and double-lobe structures in radio, and the rest-frame UV line
and continuum emission are spatially extended along the radio axis,
forming an elongated `S' or `8'-shaped structure with some knots
(Rush et al. 1997).
Rush et al. (1997) investigated the origin of the aligned continuum based on
their multi-wavelength imaging analyses of MRC 0406$-$244. 
They found that the scattering by dust in tidal features, which could be made
by a putative merger between galaxies, is favored to explain the optical depth
of each knot.
However, McCarthy (1999) suggested that the morphological appearance of
the extended nebula around MRC 0406$-$244 seen in redshifted H$\beta$ and
[O {\sc iii}] lines looks like a pair of superbubbles (see also Pentericci
et al. 2001),
which is the galactic-scale outflow driven by the collective effect of
a large number of supernova explosions of massive stars formed in a starburst
(Chevalier \& Clegg 1985; Tomisaka \& Ikeuchi 1988; see for a review
Heckman, Armus, \& Miley 1990).
In order to examine the possibility that the UV morphology may alternatively
be interpreted as a pair of superbubbles, we have made optical spectroscopy
using the Subaru 8.2 m telescope.
In this Letter, we show our new results and discuss the origin of 
the rest-frame UV and optical morphology of MRC 0406$-$244. Following Rush et al. (1997),
we adopt a Hubble constant $H_0$ = 50 km s$^{-1}$ Mpc$^{-1}$ and 
a deceleration parameter $q_0 = 0.1$ throughout this Letter.
In this cosmology, 1 arcsec corresponds to $\approx$ 10 kpc
at $z$ = 2.429.

\section{Observations}

Observations were made with the FOCAS (Faint Object Camera and 
Spectrograph: Kashikawa et al. 2000)
on the Subaru 8.2 m telescope (Kaifu 1998) during a commissioning run
of the FOCAS on 2001 February 18. The seeing condition was
0.7 arcsec.
In order to observe both line and continuum emission properties of
MRC 0406$-$244, we set the 0.8 arcsec wide slit along the major axis
of the radio structure (PA = 128$^\circ$; e.g., Rush et al. 1997).
The 300B grating together with a Y47 filter allows us to obtain an
optical spectrum between 4700 \AA ~ and 9000 \AA ~ with a 
spectroscopic resolution of 11.3 \AA ~ (the instrumental full width
at half maximum). The pixel resolution is 0.3 arcsec (3-pixel 
binning) $\times$ 5.65 \AA ~ (4-pixel binning).
The exposure time was 1800 seconds.
After subtracting the bias image in a standard manner,
flat fielding and the optical distortion corrections were applied with
the special software developed by the FOCAS team
(Yoshida et al. 2000).
Wavelength calibration and the flux calibration were made with a 
standard manner using IRAF.

\section{Results}

In the right panel of Figure 1, we show our CCD spectrum of 
MRC 0406$-$244 where the three strong emission lines, 
C {\sc iv} $\lambda$1549, He {\sc ii} $\lambda$1640, and
C {\sc iii}] $\lambda$1909, are clearly seen\footnote{Another weak emission
line O {\sc iii}]$\lambda$1663 is also seen in our spectrogram. 
The O {\sc iii}]/He {\sc ii} ratio is $\sim$ 0.1, 0.08, and 0.2 - 0.3
at 1$^{\prime\prime}$ SE, the nucleus, and 1$^{\prime\prime}$ NW, 
respectively. We do not discuss this emission line in more detail
in this Letter.}; these lines
are simply denoted as C {\sc iv}, He {\sc ii}, and C {\sc iii}].
Although this galaxy exhibits strong Ly$\alpha$ emission line
(McCarthy et al. 1996; Rush et al. 1997), our spectrum does not
cover its redshifted wavelength.
All these lines are clearly extended over 2\arcsec~ both in NW and SE
directions.
The continuum emission is found within the nuclear
$<1$\arcsec~ region, although Rush et al. (1997) detected the extended
continuum emission over $\sim 2$\arcsec~ on their high-resolution HST images.
A spectrum of the red galaxy at NW of MRC 0406$-$244 is also
recorded.
However, its redshift is unknown since it does not have any strong emission
lines in the observed wavelength coverage.
In the left hand panels, we present the spatial variations of 
emission-line velocities (top panel), full width at half maxima
(FWHM) of the emission lines corrected for the instrumental resolution
(middle panel), and the 
emission-line ratios of C {\sc iv}/He {\sc ii},
C {\sc iii}]/He {\sc ii}, and C {\sc iv}/C {\sc iii}] (bottom panel).
In all the panels, the continuum peak at $\lambda$1640 \AA ~
is set to be the origin.

The kinematical structure of the emission-line nebula is complicated across
the region.
The velocity difference between the NW and the SE sides
amounts to $\sim$ 1500 km s$^{-1}$ for the velocity curve probed
by the C {\sc iv} emission.
All emission lines are blue-shifted at most of off-nuclear regions
($\gtrsim1$\arcsec) comparing to the velocity around the nucleus. It is noted 
that the velocity curves do not show a simple symmetric structure around
the nucleus.
McCarthy et al. (1996) and Pentericci et al. (2001) found another emission-line
component with $\sim$ 1000 km s$^{-1}$ with respect to the systemic velocity
by Ly$\alpha$ emission. Our observations show that this component cannot be
probed by the three emission lines studied here.
The C {\sc iv} velocity field is more blue-shifted with respect to
that of both C {\sc iii}] and He {\sc ii} at off-nuclear regions.
The line widths of both C {\sc iv} and C {\sc iii}] emission  are
as wide as $\sim 400$ km s$^{-1}$ in FWHM at off-nuclear regions,
which is too broad to be explained by the thermal motion of star-forming nebulae within
the disk.
The nuclear region shows a narrower line width ($\sim 300$ km s$^{-1}$) in C {\sc iv}.
On the other hand, the line width of He {\sc ii} shows the different trend,
i.e., FWHM of He {\sc ii} is wider around the nucleus ($\sim 500$ km s$^{-1}$).
All these properties cannot be interpreted by a typical galactic rotation of
the host galaxy. It is noted that the different behavior of the C {\sc iv} line
from those of the other emission lines may be attributed to the large optical
depth effect because the C {\sc iv} emission line is a resonance line.

The three emission-line ratios (C {\sc iv}/He {\sc ii},
C {\sc iii}]/He {\sc ii}, and C {\sc iv}/C {\sc iii}]) are very useful to
examine excitation mechanisms of the ionized gas
(photoionization by massive stars or AGN, or shock-heating)
in high-$z$ objects\footnote{It is noted that He {\sc ii} $\lambda$1640
values in the shock+precursor models of Allen et al. (1998) are underestimated
by a factor of 1 - 6; see p. 530 of De Breuck et al. (2000).}
(e.g., Allen, Dopita, \& Tsvetanov 1998; De Breuck et al. 2000).
We find that the relative intensity of C {\sc iv} increases with increasing
radial distance (see panel c of Figure 1).
This stronger C {\sc iv} emission at large radial distance cannot
be understood in terms of photoionization by massive stars since
the ionization potential of the line (64.5 eV) is too high to 
be ionized by the ionizing radiation from massive stars.
In order to understand the excitation mechanism more clearly, we present UV
emission-line diagnostics in Figure 2;
a) C {\sc iii}]/He {\sc ii} vs. C {\sc iv}/C {\sc iii}],
b) C {\sc iv}/He {\sc ii} vs. C {\sc iv}/C {\sc iii}], and
c) C {\sc iv}/He {\sc ii} vs. C {\sc iii}]/He {\sc ii}.
The observed data points are shown together with 
model results of both AGN photoionization (Sutherland, Bicknell,
\& Dopita 1993) and shock heating (Dopita \& Sutherland 1996); see,
for the parameters adopted in the models, the upper right panel
of Figure 2.

In all diagrams, we find that both the nuclear region and the inner regions
within $r < 2$ arcsec (i.e., $r < 20$ kpc)
can be understood in terms of the AGN photoionization 
with the ionization parameter from log $U \sim -2.5$ to $-2$.
It is interesting to note that the ionization parameter increases
with increasing radial distance. Since the ionizing photon density
should decrease with the $r^{-2}$ law, it is necessary to invoke that
the nucleon density could decrease as $\propto r^a$ with $a < -2$.
An alternative idea may be that 
the shock heating is more dominated in the outer regions. 

In particular, the large departure from the
prediction of AGN photoionization can be seen in the outermost part
of the SE nebula.  There seem to be two possible 
explanations for this departure. One is the metallicity effect
because the strengths of both C {\sc iv} and C {\sc iii}
emission lines become to be larger with decreasing carbon abundance
(e.g., Vernet et al. 2001). In order to explain the observed
sudden change, it is necessary to assume that the carbon abundance decreases
suddenly by a factor of 2.5 in the outer regions. 
This value seems not surprising because some HzPRGs show evidence for such
metal-poor ionized gas (Villar-Mart\'in et al. 1997; De Breuck et al. 2000;
Overzier et al. 2001; Vernet et al. 2001).
Another idea may be the shock heating. As shown in Figure 2, 
the SE emission-line component tends to favor the shock
model with the shock velocity of $v_{\rm shock} \sim$ 200 km s$^{-1}$.
This inferred shock velocity appears consistent with the observed FWHM
of the SE nebula; see panel (b) in Figure 1.
Although it seems hard to judge which is the case for the outer part of
the SE nebula of
MRC 0406$-$244, the shock heating may contribute in part to the ionization.

\section{Discussion}

\subsection{Origin of the Figure-8 Shaped Nebulae}

Our new optical spectroscopy has shown that the spatially extended nebulae
seen in both the NW and SE directions are mainly ionized by AGN photoionization
although the SE nebula shows possible evidence for the shock heating.
If we interpret that the SE nebula contains metal-poor (i.e., $\sim 0.4 Z_\odot$)
gas photoionized by the AGN nonthermal continuum, it is necessary to
assume that the metallicity of the ionized gas shows the sudden change
at $r \sim$ 20 kpc. This sudden change may be interpreted by an idea 
that gas clouds swept up
by the superwind activity could interact with the ambient halo gas at this radius.
If we regard that the figure-8 shaped morphology of the SE and NW nebulae
is attributed to wound tidal features (Rush et al 1997), 
it seems hard to explain the sudden change in metallicity. 
It is therefore considered that the figure-8 shaped morphology can be interpreted
as the relic of a pair of superbubbles.

The origin of UV continuum of the figure-8 shaped structure is not clear.
Since the majority of the ionization can be attributed to the AGN photoionization,
massive stars could not be the major UV continuum source. Stars later than
B-type stars could contribute to the UV continuum if they were made in the
shock heated gas clouds. 
An alternative idea is the scattering of
an anisotropic continuum radiation from a central engine of AGN
(e.g., Cimatti et al. 1998 and references therein).
However, no spectropolarimetry has not yet been done for
MRC 0406$-$244 and thus we cannot conclude which is the case.
It will be important to carry out optical spectropolarimetry to investigate
this issue in future.

\subsection{The Origin of the Alignment Effect of MRC 0406$-$244}

Here we consider a possible origin of the alignment effect of MRC 0406$-$244.
We attribute the figure-8 shaped morphology of this galaxy to a pair of
superbubbles. 
Since MRC 0406$-$244 is associated with a bright host galaxy
(i.e., $M_K \sim -29$ or $M_V \sim -25$), the host galaxy may be
a typical massive galaxy with a mass of $\sim 10^{12} M_\odot$.
If this is the case, it takes $\sim 7 \times 10^8$ yr to the onset
of the superbubble from the host galaxy potential (Arimoto \& Yoshii 1987).
This age appears consistent with that estimated by Eales \&
Rawlings (1993; see also McCarthy et al. 1992).

On the other hand, the age of the radio jet ($t_{\rm jet}$) in MRC 0406$-$244 
is much shorter than the above timescale because the projected
distances of the NW and SE radio lobes, 54.5 kpc and 29.5 kpc,
gives nominally $t_{\rm jet} \sim 2 \times 10^6$ yr and 
$\sim 1 \times 10^6$ yr, respectively, given the jet velocity
of $v_{\rm jet} \sim 0.1c$ (e.g., Alexander \& Leahy 1987).
Therefore, it seems reasonable to consider that the radio jet
activity is an independent phenomenon from the starburst
and the subsequent superwind activity. This seems to make sense
because the onset of the radio-jet activity needs the presence 
of a supermassive black hole and the gas accretion onto it
(e.g., Rees 1984), both of which are different things from 
the starburst activity. 

Here a question arises as why the radio jet axis is aligned
to the superwind direction since the radio-jet activity and
the starburst/superwind activity are different physical processes
and their timescales are also significantly different.
The superwind could blow as a bipolar wind, which is often observed in many
superwind galaxies in the local universe (see Heckman et al. 1990),
since it is likely that the gas
in the host galaxy is distributed with a disk-like configuration even for 
the young host galaxy of MRC 0406$-$244.
On the other hand, the radio jet is expelled to the two directions
perpendicular to the accretion plasma disk.
Therefore, we can explain the alignment effect if
the accretion disk is nearly co-planer to the host disk.
Another example to which this scenario
is applicable may be B3 0731+438 at $z$ = 2.43 studied by
Motohara et al. (2000), which also shows the alignment effect with
the biconical nebula.

\subsection{Other Implications}

MRC 0406$-$244 tells us that the intense starburst and subsequent 
superwind activity may be important even in HzPRGs. 
In particular, the formation of shell-like structures driven
by the superwind activity gives rise to some important implication.
For example, if low-mass stars are also
formed in blobs, the blobs could evolve either to globular clusters
(Taniguchi, Trentham, \& Ikeuchi 1999) or to dwarf galaxies
(e.g., Mori, Yoshii, \& Nomoto 1999). Another example of such
superwind/superbubble-driven formation of shell-like structures 
may be LRG J0239-0134 at $z \sim 1$ (Taniguchi \& Murayama 2001).
Since high-redshift superwinds could contribute to the chemical
enrichment of the intergalactic medium (e.g., Heckman et al. 1990;
Taniguchi, \& Shioya 2001), it will be important to investigate
HzPRGs systematically.

\vspace{0.5cm}

We would like to the staff of the Subaru Telescope Office.
We also thank Fumihide Iwamuro and Carlos De Breuck for useful discussion on
HzPRGs and the referee, Patrick McCarthy for useful comments
and suggestions.
This work was financially supported in part by
the Ministry of Education, Science, and Culture
(Nos. 10044052, and 10304013).



\figcaption{
a) The velocity curves obtained using the C {\sc iv}, C {\sc iii}],
and He {\sc ii} emission lines.
b) The spatial variations of FWHM of the C {\sc iv}, C {\sc iii}],
and He {\sc ii} emission lines.
c) The spatial variations of the following line ratios:
C {\sc iv}/He {\sc ii},
C {\sc iii}]/He {\sc ii}, and C {\sc iv}/C {\sc iii}].
d) The optical spectrogram of MRC 0406$-$422.
\label{fig1}}

\figcaption{
UV emission-line diagnostics diagrams;
a) C {\sc iii}]/He {\sc ii} vs. C {\sc iv}/C {\sc iii}],
b) C {\sc iv}/He {\sc ii} vs. C {\sc iv}/C {\sc iii}],
and c) C {\sc iv}/He {\sc ii} vs. C {\sc iii}]/He {\sc ii}].
Details of the model results are given in the upper right panel.
\label{fig2}}

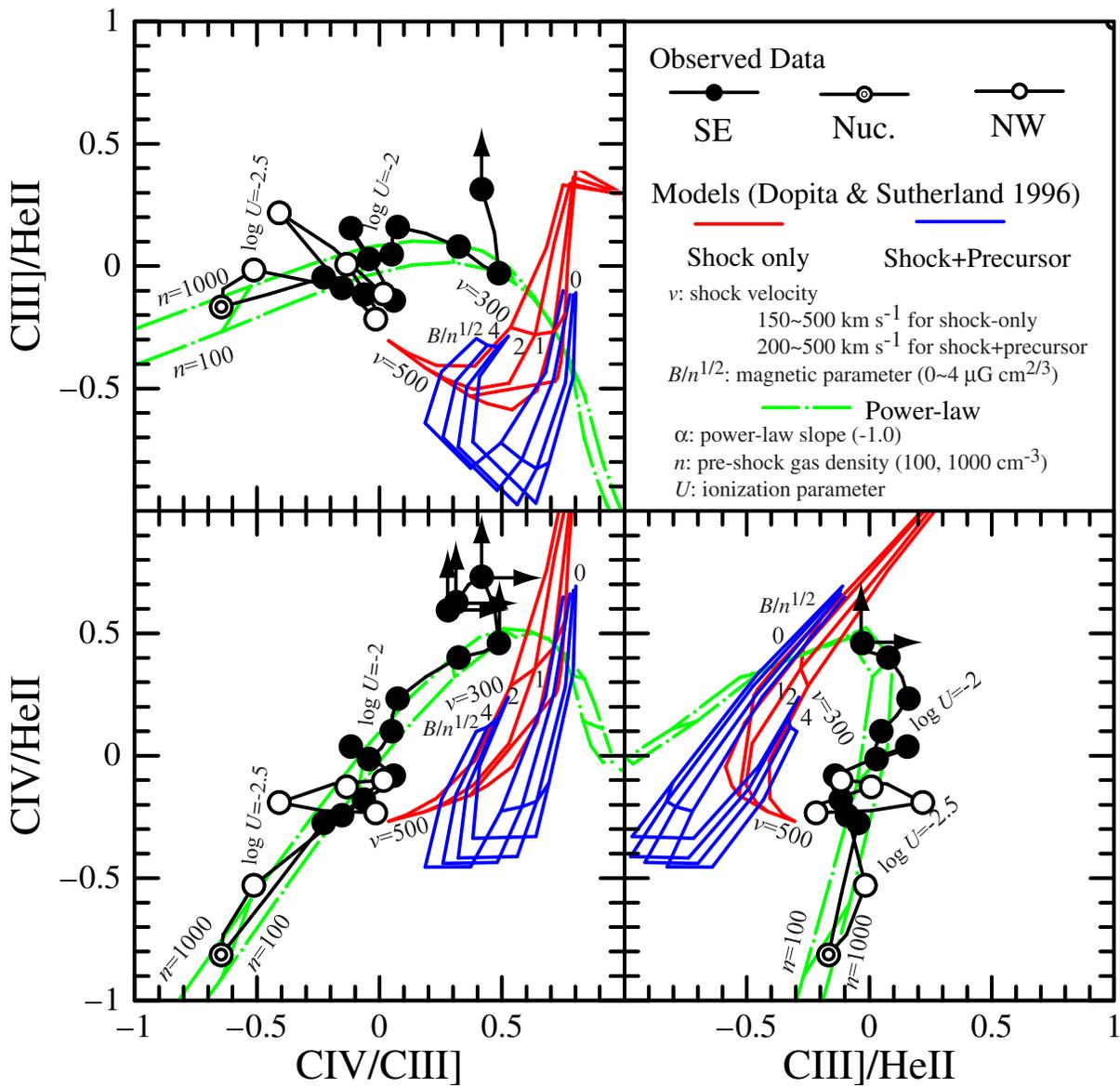
\end{document}